\documentclass[conference]{IEEEtran}
\IEEEoverridecommandlockouts
\usepackage{cite}
\usepackage{amsmath,amssymb,amsfonts}
\usepackage{algorithmic}
\usepackage{graphicx}
\usepackage{textcomp}
\newcommand{\model}{DCLab}
\def\BibTeX{{\rm B\kern-.05em{\sc i\kern-.025em b}\kern-.08em
    T\kern-.1667em\lower.7ex\hbox{E}\kern-.125emX}}
\begin{document}

\title{\model: A Web-based System for Digital\\Logic Experiment Teaching}

\author{\IEEEauthorblockN{Yuhui Ding, Shanshan Li, Jinghan Liu and Xiaojun Wu}
\IEEEauthorblockA{Department of Computer Science and Technology \\
Tsinghua University \\
Beijing, China \\
dingyh15@mails.tsinghua.edu.cn, lishanshan@tsinghua.edu.cn, LiuJh@lib.tsinghua.edu.cn, xjwu@tsinghua.edu.cn}
}

\maketitle

\begin{abstract}
This Research-to-Practice Work in Progress paper presents \model, a web-based system for conducting digital logic experiments online, to improve both the effectiveness and the efficiency of digital logic experiment teaching. \model \space covers all experimental contents required by traditional digital logic experiment classes. It allows the students to draw circuit diagrams or to write VHDL code to design their own circuits, and it provides complete simulation functions. \model \space records the progress of each student and makes it convenient for a student to review the history of his practice. Furthermore, for the instructors, they are able to post homework assignments on \model, and the system will automatically add homework projects to the students' home pages. Statistics about how the students perform on the homework will be displayed to the instructors, which may help them develop more effective courses. We have tested \model \space among students in the digital logic course, and the results have confirmed its validity.
\end{abstract}


\section{Introduction}
Digital Logic is a dingyh fundamental course in the computer science curriculum, and its experimental study plays an important role in connecting theory and practice\cite{yan2007innovation}\cite{adamo2009innovative}, which helps students consolidate what they have learned from textbooks. Specifically, there are two types of digital logic experiment\cite{shanshan2017digital}. One is to require the students to build digital circuits manually, connecting circuit elements such as chips with wires, and use instruments for debugging. This type of experiment is indispensable in helping the students get a deep understanding of the digital logic theory and developing their hands-on skills. The other is to use programmable devices, and the students write the hardware description language (HDL) code to design circuits. In this way the students are able to keep away from tedious manual operations, and they can implement complex circuits more easily.

For classical approaches, a detailed guideline and some hardware equipments are distributed before the experimental task, and the students are required to finish design and make measurements on the given devices. Those hardware equipments are often somewhat unwieldy, leading to inconvenience for students who have to bring them to the laboratory. For the instructors, they need to examine the experiment results one by one, and they are sometimes bothered by the high damage rate of equipments. 

To improve both the effectiveness and the efficiency of digital logic experiment teaching, we have developed \model, an integrated web-based system. \model \space is mainly composed of four important modules: Graphical Circuit Module, VHDL Module, Simulation Module and Management Module. The Graphical Circuit Module enables students to place different circuit components on canvas and then assemble a circuit virtually. The VHDL Module allows students to edit circuit design files and stimulation files using VHDL. The Simulation Module is responsible for performing simulation and showing simulation waveforms. The Management Module manages users' projects and allows the instructors to assign homework.

To sum up, the main contributions of our proposed system are listed as follows:
\begin{itemize}
\item Based on the web, \model \space allows students to carry out digital logic experiments virtually, involving functions required for a complete experiment process such as compilation, simulation etc. It saves time and labor for both students and instructors.
\item \model \space supports two types of digital logic experiment: graphical circuit building and VHDL editing, taking advantages of both.
\item Even though conducting experiments online may mean less face-to-face time, \model \space keeps good interaction between students and instructors. It allows instructors to assign homework and track the students' progress, so that they can give directions accordingly.
\end{itemize}

The rest of the paper is organized as follows: We briefly review related work in Section \uppercase\expandafter{\romannumeral2}. In Section \uppercase\expandafter{\romannumeral3} we introduce the architecture of \model \space in detail. In Section \uppercase\expandafter{\romannumeral4} we present experiment results and we conclude in Section \uppercase\expandafter{\romannumeral5}.

\section{Related Work}
Considering the importance of digital logic experiment teaching, many educators have been focused on how to make it more effective and how to promote the students' innovation ability\cite{farook2011computer}. Li et al.\cite{shanshan2017training} set up two different experimental courses according to students' interests and knowledge levels, and they design special hardware equipments which have various interfaces. Even though the particular hardware equipment supports many experimental contents and meets the requirements of the educators, it is rather inconvenient for the students to carry equipments with them, and they are very likely to be confused sometimes by too many wires which twine in the limited space.

To address the aforementioned problems, Yang et al.\cite{yang2007web} and Kim et al.\cite{kim2009web} both describe the design of a virtual laboratory for digital circuit experiments. The system in \cite{yang2007web} only supports drawing wires through drag and drop on pins of circuit elements on a fixed board, which restricts the students' innovative ideas, and it doesn't allow students to design circuits using HDL. What's more, the students are only able to interact with the server that works out simulation results, thus lacking interactive learning between students and instructors. The system in \cite{kim2009web} adds flash animation to explain the key concepts of the experiment, and it will give some multiple questions to users after virtual experiments to assess the learning effect. Nonetheless, it cannot support programmable experiments, and it still faces the problem of little interaction between the students and the instructors.

\section{System Architecture}
\begin{figure}[h]
\centering
\includegraphics[width=0.5\textwidth]{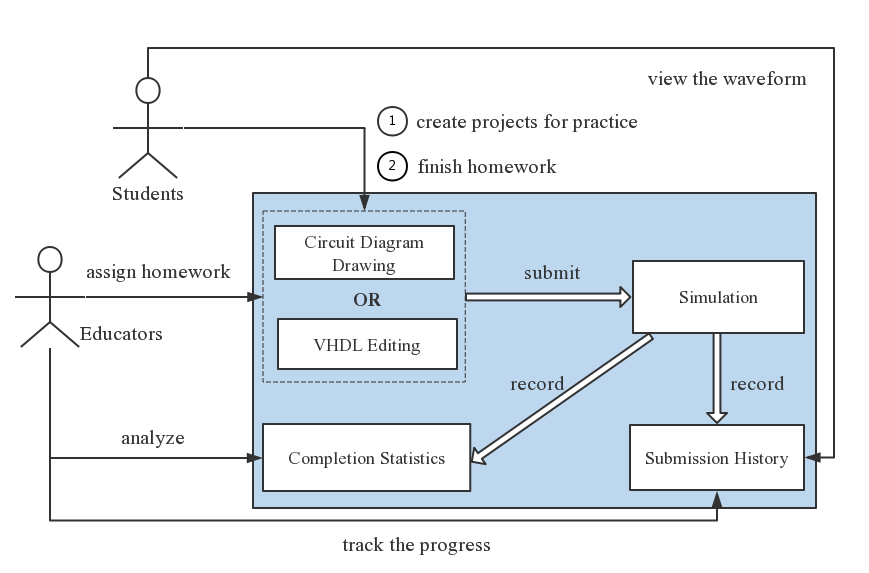}
\caption{System architecture of \model}
\label{fig:architecture}
\end{figure}

To meet the requirements of the digital logic course and fully support students' experiment practice, \model \space has integrated all activities a student in the traditional experiment course may get involved in, such as design, simulation etc. The whole architecture of \model \space is illustrated in Figure \ref{fig:architecture}. There are two ways to use the system in practice. For one thing, the students can create their own projects, drawing circuit diagrams or editing VHDL files, and see the simulation results. For another, the students may be required to finish experimental tasks as homework assignments, and their homework performance will be shown to the instructors. The functions of \model \space are implemented by several different modules, which will be demonstrated in detail next.

\subsection{Graphical Circuit Module}
The Graphical Circuit Module provides a virtual environment where students can place different circuit elements on canvas and use wires to connect them through merely a few clicks. We have implemented all circuit elements which are required for the digital logic course, including all chips of the 74 series, seven segments LED displays, digital decoders etc. In building the circuit, students can drag and drop circuit elements freely on canvas, change their sizes to some extent and use shortcuts conveniently for cancel or deletion. What's more, \model \space is intelligible enough to discern some common operation mistakes. For instance, warnings will be given when a short circuit is detected, and it is not permitted to draw a wire between two output ports.

%\begin{figure}[h]
%\centering
%\includegraphics[width=0.5\textwidth]{image/graphic.png}
%\caption{The Graphical Circuit Setup Module interface}
%\label{fig:graphic}
%\end{figure}

\begin{figure}[h]
\centering
\includegraphics[width=0.5\textwidth]{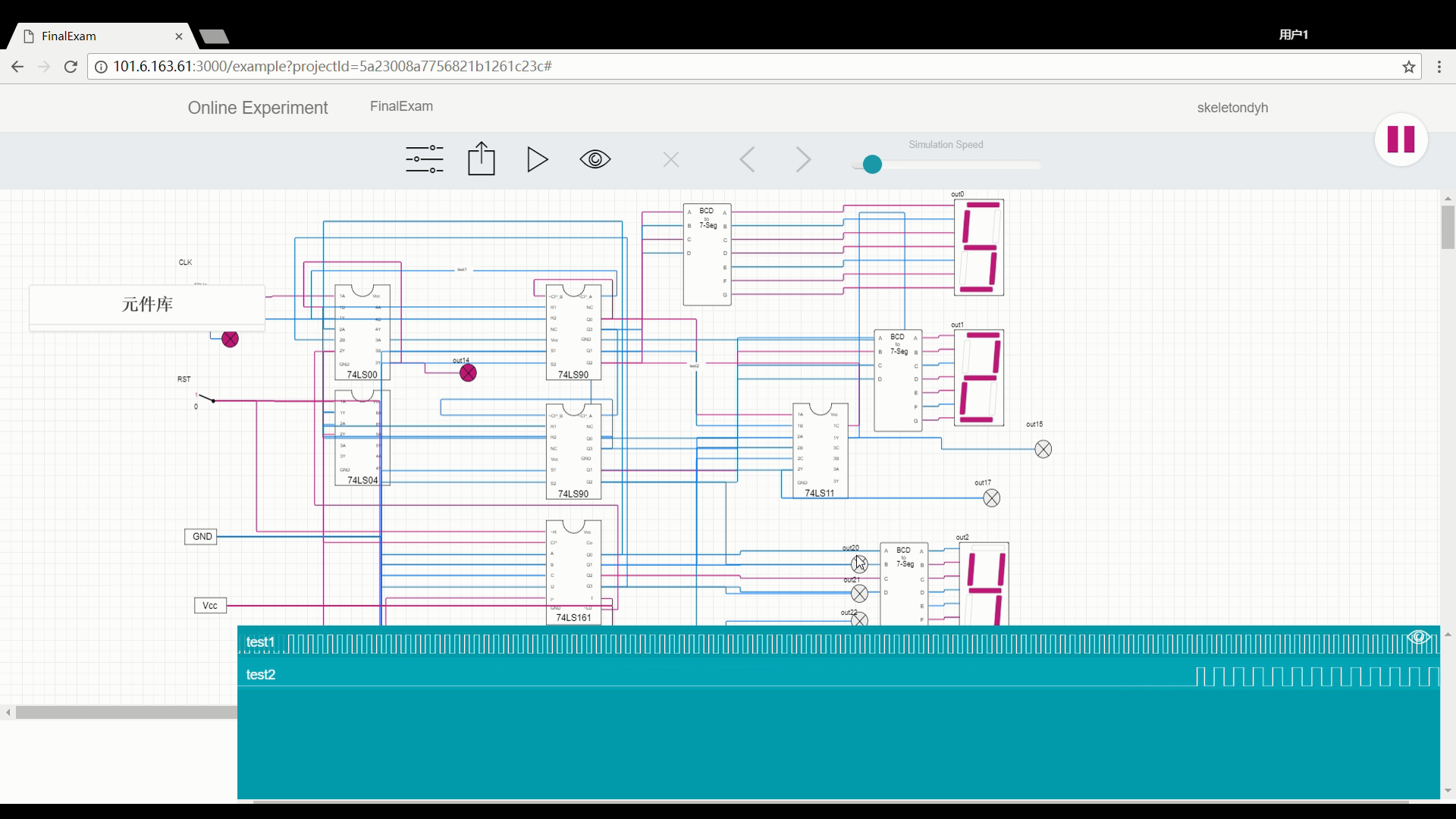}
\caption{The simulation of a graphical circuit and the probes}
\label{fig:simulation}
\end{figure}

One of the most amazing features of the Graphical Circuit Module is that it provides two simulation methods. One method is to simulate the graphical circuit directly in the browser, without sending any file to the background server. Connecting some input stimulation such as a 50Hz clock frequency to the circuit, students can visually see whether the graphical circuit works correctly. Furthermore, in the simulation process, students can place probes to detect the status of different positions in the circuit, and the waveforms of the viewed points changing with time will be displayed at the bottom of the interface, which is illustrated in Figure \ref{fig:simulation}. In this way, students are able to inspect their circuits intuitively and get their ability of debugging improved by using probes. Also, this kind of simulation helps save the computing resources of the server.

\begin{figure}[h]
\centering
\includegraphics[width=0.5\textwidth]{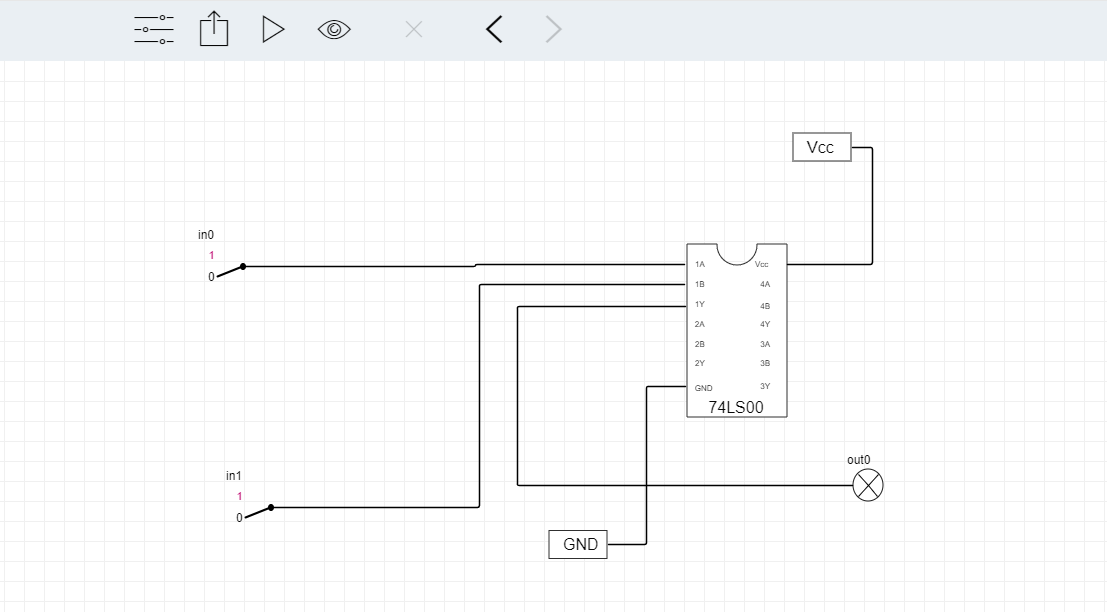}
\includegraphics[width=0.5\textwidth]{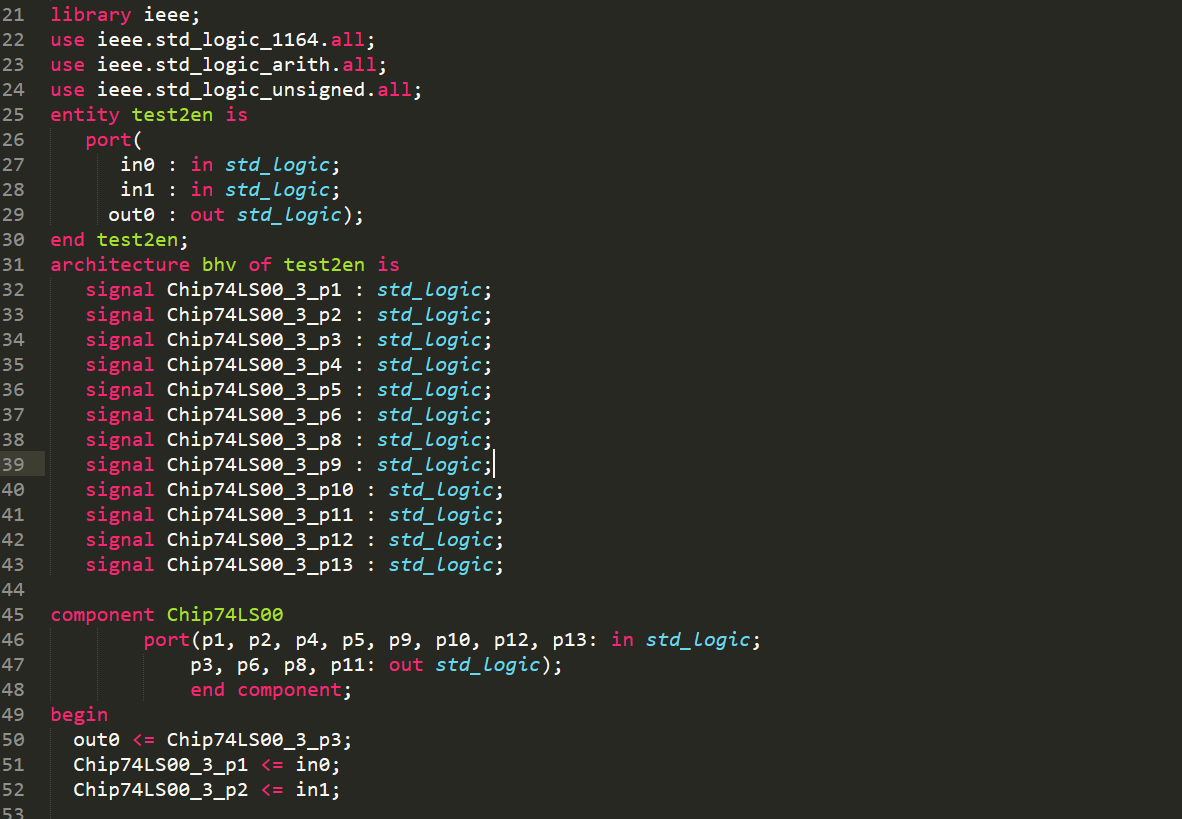}
\caption{A 74LS00 chip and its corresponding VHDL code}
\label{fig:code}
\end{figure}

\begin{figure}[htbp]
\centering
\includegraphics[width=0.5\textwidth]{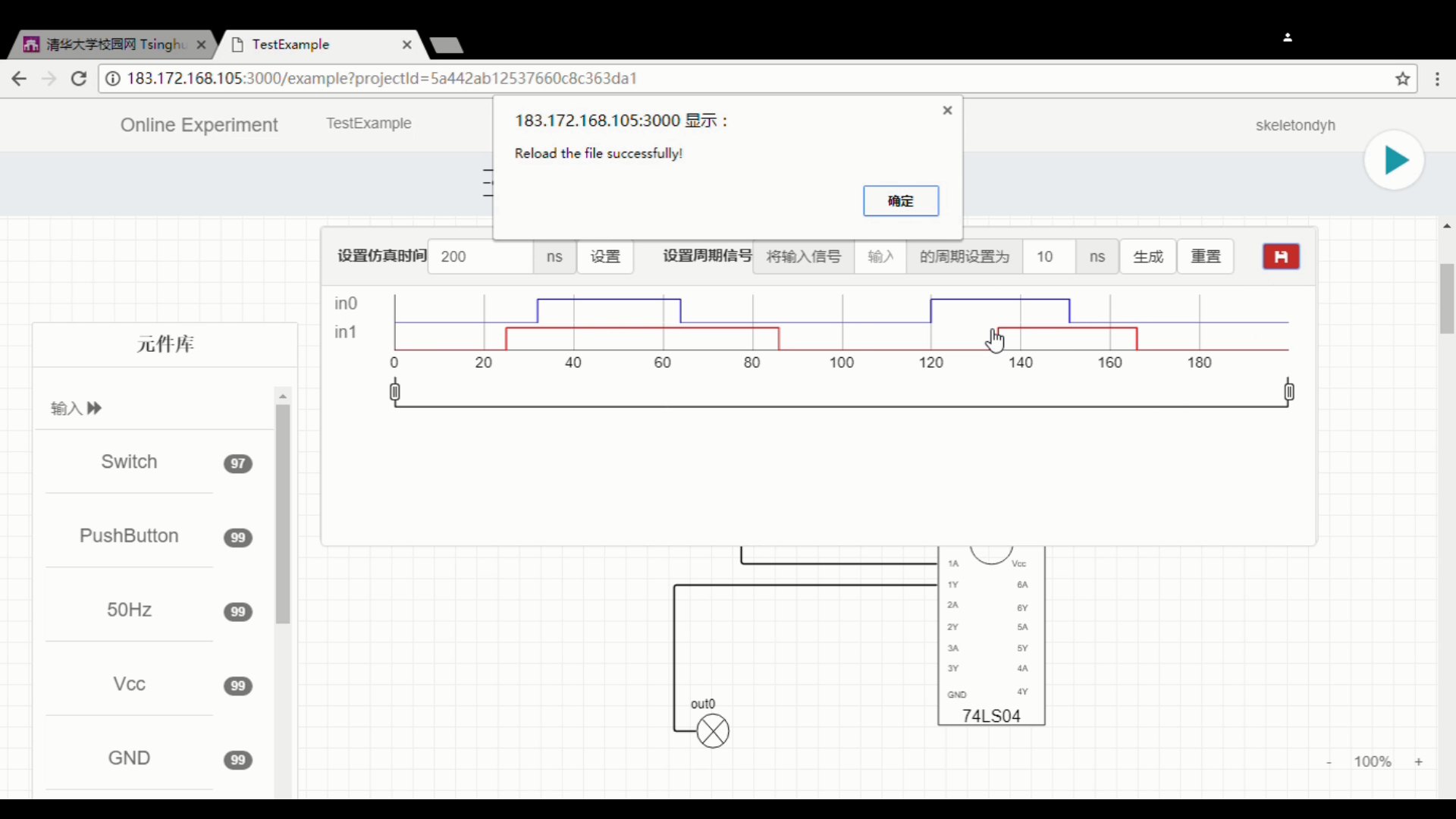}
\caption{Stimulation signal editor}
\label{fig:signal}
\end{figure}

The other simulation method gives more detailed results. The students can submit the circuit and the stimulation signals to the background server, and when the data are received by the server they are automatically converted to VHDL files to perform simulation steps using ModelSim. Figure \ref{fig:code} shows an example circuit and its corresponding VHDL code. \model \space provides a graphical user interface to edit the stimulation signals manually, as is shown in Figure \ref{fig:signal}. Entering the stimulation signal editor, the system will automatically read labels of the input ports and display them in the interface, then the students can either draw waveforms or set some parameters in the text fields to edit the input signals.

\subsection{VHDL Module}
The VHDL Module has integrated an open source framework to support students in writing VHDL code directly to design the circuit and define input signals. The VHDL Module acts like other integrated development environments (IDE), providing sophisticated editing features such as color syntax highlighting, keyboard macros, code completion etc. These features help the students learn VHDL quickly and improve the efficiency of writing code. The compiler is able to work with multiple source files, and if any error occurs in the compilation, the debugger may suggest what type it is and what line it is on.

%\begin{figure}[h]
%\centering
%\includegraphics[width=0.5\textwidth]{image/circuit.PNG}
%\includegraphics[width=0.5\textwidth]{image/code.PNG}
%\caption{A 74LS00 chip and its corresponding VHDL code}
%\label{fig:signal}
%\end{figure}

\subsection{Simulation Module}
\begin{figure}[h]
\centering
\includegraphics[width=0.5\textwidth]{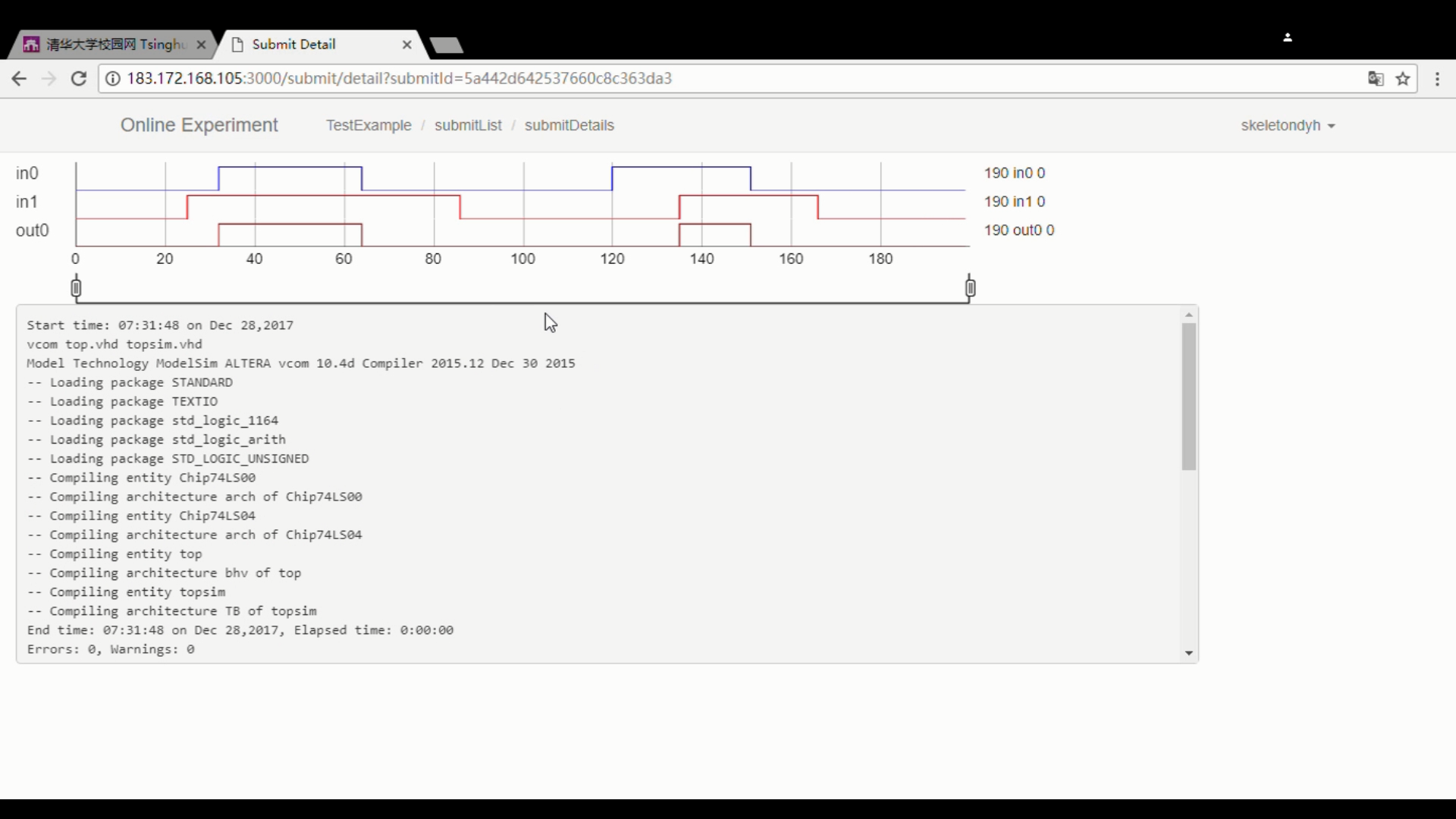}
\caption{The simulation waveforms of a submission}
\label{fig:result}
\end{figure}

The Simulation Module employs ModelSim to conduct simulation of VHDL files describing circuit design and input signals. The background server will execute certain ModelSim commands after it receives submitted files and will then return simulation waveforms and logs to be displayed to the user. Figure \ref{fig:result} shows the result of a submission.

\subsection{Management Module}
The Management Module is based on a database which contains all user profiles, projects' metadata and submission records. It is of great importance in making \model \space an interactive system.

%\begin{figure}[h]
%\centering
%\includegraphics[width=0.5\textwidth]{image/index.png}
%\caption{The index page for students}
%\label{fig:index}
%\end{figure}

On a student entering his home page, he will see four columns: Attention, Homework, Playground and Example. The Attention column lists notices with their dates of issue and authors, so that the students can read important experiment instructions from the instructors promptly. The Homework column contains experimental projects which the students are required to finish and submit as homework. The Management Module will automatically add an empty project to every student's Homework column once a new homework assignment is posted. In the Playground column, the students are free to create their own projects. While the type of a homework project (graphical circuit or VHDL code) is often specified by the instructor, a student can decide how to design his circuit independently for a self-created project, and he is encouraged to design some innovative circuits to achieve some particular functions. The Management Module records every submission of a self-created project in the database and keeps related files, so that a student can review the submission history of a certain project at any time. Each item of the submission history contains simulation waveforms (if compiled correctly) and compilation logs. Students' projects and submission histories can also be accessed by the instructors, which enable them to track the students' progress and give guidance accordingly. Projects in the Example column are edited by the instructors as paradigms or homework answers. The students do not have permissions to modify an example project, but they can conduct a simulation analysis and verify its correctness. 

%\begin{figure}[h]
%\centering
%\includegraphics[width=0.5\textwidth]{image/teacher.png}
%\caption{The user interface for the educators}
%\label{fig:teacher}
%\end{figure}

On the side of instructors, the Management Module allows them to broadcast notices, edit examples and assign homework. Example projects serve as paradigms or homework answers which the students may refer to. Editing examples is almost the same as what students would do to create their own projects. Once an example project is built, instructors could decide whether to set it visible to all students. To assign homework, the instructors are required to specify an example project as the possible answer and edit several different input stimulations, using the graphical signal editor, as test points. After a student has submitted his homework, \model \space will automatically complete compilation and simulation analysis, and will then check whether the obtained simulation waveforms are consistent with those of the example at all test points. The submission is graded according to how many test points it has passed. 

Inspired by \cite{minaei2003predicting}, in order to make the instructors understand how well the students perform on the homework assignment, the Management Module records the following features for each individual student:
%\begin{itemize}
%\item The ratio of students who have already submitted homework at least once
%\item The distribution of numbers of submissions for the homework.
%\item The distribution of scores of students who have already submitted homework (The score of a student is the highest score among all his submissions before the homework deadline.)
%\end{itemize}
\begin{itemize}
\item The number of submissions for the homework assignment
\item The time of each submission
\item The score of each submission
\item The score of the student for the homework assignment (the highest score among all his submissions before the homework deadline)
\end{itemize}

Through these features the instructors are able to know an individual student's progress promptly and could then help him solve some possible problems. What's more, some metrics which can reflex the students' performance in general, such as the ratio of students who have already submitted the homework at least once, can be drawn from individual features of all students. These metrics are calculated automatically and displayed to the instructors through diagrams, and they can help the instructors develop the homework more effectively.
%Besides, the educators are able to view the status and the submission history of any student project. For assigned homework, the system will periodically update the statistics about students' completion before the due date and display them visually on the interface, including the ratio of students who have submitted the homework, the distribution of scores etc. The final score of a student will be the highest score among his submissions by the homework deadline. In this way, the educators can know the progress of the students promptly, and thus modify the content or schedule of the experiment teaching accordingly.

\section{Experiment Results}
In order to show the validity of \model, we have tested it within a group of 31 second year undergraduates in the digital logic course. The students were instructed to design a simple counter circuit which could count from 0 to 59 in decimal and display the digits using the seven segments LED displays. They were allowed either to assemble a graphical circuit or to write VHDL code. We collected data three days after the homework was assigned without informing the students.

\begin{figure}[h]
\centering
\includegraphics[width=0.5\textwidth]{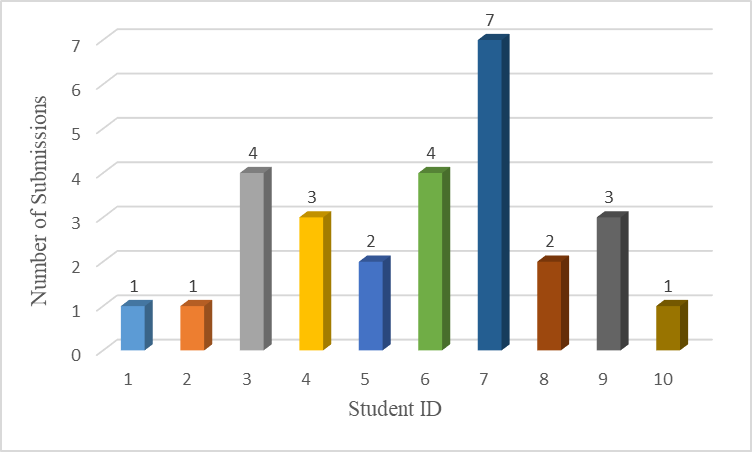}
\caption{How many submissions the students have made}
\label{fig:submission}
\end{figure}

Among the 31 students, 17 have submitted the homework at least once, 10 of which have got the correct solution. Considering students' different levels, those who have got the correct answer may have different numbers of tries for doing homework. Figure \ref{fig:submission} illustrates how many submissions the students have made before they get the right answer, which verifies our assumption. From Figure \ref{fig:submission}, we see that while student 1, 2 and 10 have given the right solution in their first submission, student 7 may have met more problems and has tried 7 times. It would cost him much more time and labor if he debugged his circuit in traditional experiment courses. For the instructors, such information provided by \model \space allows them to focus on certain particular students and help them accordingly.

%\begin{figure}[h]
%\centering
%\includegraphics[width=0.5\textwidth]{image/hiahia.png}
%\caption{Distribution of submissions with respect to time}
%\label{fig:time}
%\end{figure}

\begin{figure}[h]
\centering
\includegraphics[width=0.5\textwidth]{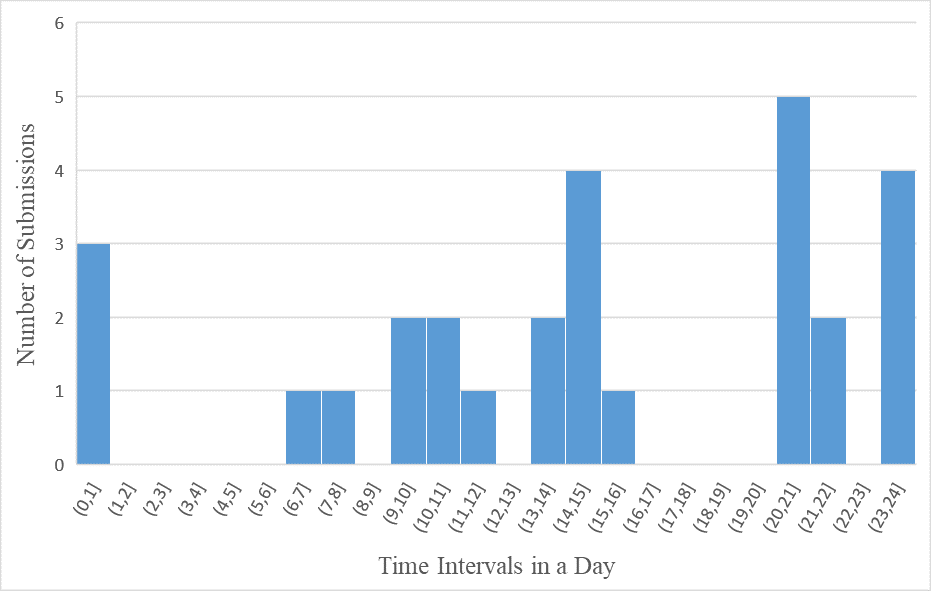}
\caption{Distribution of submissions with respect to time}
\label{fig:time}
\end{figure}

We have also looked over when the students made their submissions. Figure \ref{fig:time} shows the distribution of submissions among the 24 one-hour intervals in a day, which demonstrates that \model \space provides a large flexibility for the students to choose suitable time to finish their experimental tasks.

%\begin{figure}[h]
%\centering
%\includegraphics[width=0.5\textwidth]{image/xixi.png}
%\caption{Number of submissions the students have made}
%\label{fig:submission}
%\end{figure}

\section{Conclusion and future work}
In this paper, we propose \model, a web-based system for digital logic experiment teaching. \model \space combines two types of digital logic experiment and meets the requirements of the digital logic experiment course. Besides, \model \space keeps good interaction between students and instructors, which makes it suitable for educational practice. Experiment results have shown that \model \space can improve both the effectiveness and the efficiency of digital logic experiment teaching. 
%We have developed a web-based system for digital logic experiment teaching, which combines two types of digital logic experiment and meets the requirements of the digital logic course. This system improves the efficiency of both students and educators, and in the meantime maintains good interactions between them. Experiment results have demonstrated the validity of our system. 

For future work, we will employ \model \space in courses with a larger number of students, and try to adapt it to the needs of students and instructors. 
%\begin{figure}
%\centering
%\includegraphics[width=0.5\textwidth]{image/history.png}
%\caption{The submission history of a student for homework}
%\label{fig:history}
%\end{figure}

\bibliographystyle{IEEEtran}
\bibliography{IEEEabrv,reference}

\begin{thebibliography}{1}
\providecommand{\url}[1]{#1}
\csname url@samestyle\endcsname
\providecommand{\newblock}{\relax}
\providecommand{\bibinfo}[2]{#2}
\providecommand{\BIBentrySTDinterwordspacing}{\spaceskip=0pt\relax}
\providecommand{\BIBentryALTinterwordstretchfactor}{4}
\providecommand{\BIBentryALTinterwordspacing}{\spaceskip=\fontdimen2\font plus
\BIBentryALTinterwordstretchfactor\fontdimen3\font minus
  \fontdimen4\font\relax}
\providecommand{\BIBforeignlanguage}[2]{{%
\expandafter\ifx\csname l@#1\endcsname\relax
\typeout{** WARNING: IEEEtran.bst: No hyphenation pattern has been}%
\typeout{** loaded for the language `#1'. Using the pattern for}%
\typeout{** the default language instead.}%
\else
\language=\csname l@#1\endcsname
\fi
#2}}
\providecommand{\BIBdecl}{\relax}
\BIBdecl

\bibitem{yan2007innovation}
X.~Yan-xia, ``Innovation and research of experimental teaching for digital
  electric circuit [j],'' \emph{Research and Exploration in Laboratory},
  vol.~2, pp. 84--86, 2007.

\bibitem{adamo2009innovative}
O.~B. Adamo, P.~Guturu, and M.~R. Varanasi, ``An innovative method of teaching
  digital system design in an undergraduate electrical and computer engineering
  curriculum,'' in \emph{Microelectronic Systems Education, 2009. MSE'09. IEEE
  International Conference on}.\hskip 1em plus 0.5em minus 0.4em\relax IEEE,
  2009, pp. 25--28.

\bibitem{shanshan2017digital}
L.~Shanshan and Y.~Shiqiang, ``Digital logic experiment teaching based on
  experimental platform,'' in \emph{Computer Science and Education (ICCSE),
  2017 12th International Conference on}.\hskip 1em plus 0.5em minus
  0.4em\relax IEEE, 2017, pp. 33--37.

\bibitem{farook2011computer}
O.~Farook, C.~R. Sekhar, J.~P. Agrawal, E.~Bouktache, A.~Ahmed, and
  H.~Moghbelli, ``Computer engineering technology program-a curriculum
  innovation initiative,'' in \emph{American Society for Engineering
  Education}.\hskip 1em plus 0.5em minus 0.4em\relax American Society for
  Engineering Education, 2011.

\bibitem{shanshan2017training}
L.~Shanshan, W.~Xiaojun, and Q.~Chengbin, ``Training students' practical and
  innovation ability in hardware experiment,'' in \emph{Frontiers in Education
  Conference (FIE)}.\hskip 1em plus 0.5em minus 0.4em\relax IEEE, 2017, pp.
  1--5.

\bibitem{yang2007web}
O.~Yang, Y.~Shiping, D.~Yabo, and Z.~Miaoliang, ``Web-based interactive virtual
  laboratory system for digital circuit experiment,'' in \emph{Innovations in
  E-learning, Instruction Technology, Assessment, and Engineering
  Education}.\hskip 1em plus 0.5em minus 0.4em\relax Springer, 2007, pp.
  305--309.

\bibitem{kim2009web}
D.~Kim, K.~Choi, C.~Jeon, J.~Lim, S.~Kim, S.~Seo, and J.~Yoo, ``A web-based
  virtual laboratory system for electronic and digital circuits experiments,''
  in \emph{International Conference on Hybrid Learning and Education}.\hskip
  1em plus 0.5em minus 0.4em\relax Springer, 2009, pp. 77--88.

\bibitem{minaei2003predicting}
B.~Minaei-Bidgoli, D.~A. Kashy, G.~Kortemeyer, and W.~F. Punch, ``Predicting
  student performance: an application of data mining methods with an
  educational web-based system,'' in \emph{Frontiers in education, 2003. FIE
  2003 33rd annual}, vol.~1.\hskip 1em plus 0.5em minus 0.4em\relax IEEE, 2003,
  pp. T2A--13.

\end{thebibliography}

\end{document}